\documentclass[12pt,showkeys,prab,superscriptaddress,floatfix,nofootinbib,longbibliography]{revtex4-2}

\usepackage{graphicx}
\usepackage{color}
\usepackage{amsfonts}
\usepackage{bm}
\usepackage{xspace}

\newcommand {\bfr} {{\bf r}}
\newcommand {\bfv} {{\bf v}}

\newcommand {\bfB} {{\bf B}}
\newcommand {\bfE} {{\bf E}}
\newcommand {\bfF} {{\bf F}}

\renewcommand {\d} {{\rm d}}
\newcommand {\ee}{{\rm e}}

\newcommand {\E} {\varepsilon}

\newcommand {\om} {\omega}
\newcommand {\Om} {\Omega}

\newcommand{\MBNExplorer} {\textsc{MBN Explorer}\xspace}
\newcommand{\MBNStudio}{\textsc{MBN Studio}\xspace}

\newcommand{\calA} {{\cal A}}
\newcommand{\Lp}{L_{\rm p}}
\newcommand{\Ld}{L_{\rm d}}
\newcommand{\Lr}{L_{\rm r}}

\begin{document}

\title{Atomistic modeling of the channeling process with and without
         account for ionising collisions: A comparative study}

\author{Gennady B. Sushko}
\affiliation{MBN Research Center, Altenh\"{o}ferallee 3, 60438 Frankfurt am Main, Germany}

\author{Andrei V. Korol}
\email[]{korol@mbnexplorer.com}
\affiliation{MBN Research Center, Altenh\"{o}ferallee 3, 60438 Frankfurt am Main, Germany}

\author{Andrey V. Solov'yov}
\email[]{solovyov@mbnresearch.com}
\affiliation{MBN Research Center, Altenh\"{o}ferallee 3, 60438 Frankfurt am Main, Germany}

\begin{abstract}
This paper presents a quantitative analysis of the impact of inelastic
collisions of ultra-relativistic electrons and positrons, passing through oriented
crystalline targets, on the channeling efficiency and on the intensity  of the
channeling radiation.
The analysis is based on the numerical simulations of the channeling process
performed using the \MBNExplorer software package.
The ionising collisions, being random, fast and local events, are
incorporated into the classical molecular dynamics framework according
to their probabilities.
This methodology is outlined in the paper.
The case studies presented refer to electrons with energy $\E$ ranging from 270
to 1500 MeV and positrons with $\E=530$ MeV incident on thick (up to 1 mm)
single diamond, silicon and germanium crystals oriented along the (110) and
(111) planar directions.
In order to elucidate the role of the ionising collisions,
the simulations were performed with and without account for the
ionising collisions.
The case studies presented demonstrate that both approaches yield highly
similar results for the electrons.
For the positrons, the ionising collisions reduce significantly the
channeling efficiency.
However, it has been observed that this effect does not result in
a corresponding change in the radiation intensity.
\end{abstract}

\maketitle

\section{Introduction  \label{Introduction}}

Interaction of high-energy charged particles with crystals
is sensitive to the orientation of the incoming beam with respect to
to the main crystallographic directions (planar and axial) of the target.
Projectiles, incident at small angles to the crystal planes
(or axes) experience specific channeling motion
following the planar (axial) direction experiencing collective action
of the electrostatic fields of the lattice atoms \cite{Lindhard}.

The study of the channeling process of ultra-relativistic particles in
oriented crystals 
has emerged as
a wide-ranging field of research \cite{BiryukovChesnokovKotovBook,Uggerhoj:RPM_v77_p1131_2005,%
ChannelingBook2014,CLS-book_2022}.
Various applications have been suggested
including beam steering
\cite{MazzolariEtAl:PRL_v112_135503_2014,MazzolariEtAl:EPJC_v78_p720_2018,
WienandsEtAl:PRL_v114_074801_2015},
collimation \cite{Scandale_EtAl-PRB_v692_p78_2010},
focusing \cite{Scandale_EtAl-NIMB_v446_p15_2019}, and
extraction \cite{BiryukovChesnokovKotovBook}.

Oriented crystals of different geometries (single, bent, periodically bent)
exposed to the beams of ultra-relativistic electrons and positrons can
potentially serve as novel intensive gamma-ray crystal-based light sources
(CLS) operating in the MeV-GeV photon energy range
\cite{CLS-book_2022,SushkoKorolSolovyov:EPJD_v76_166_2022,%
KorolSolovyov:NIMB_v537_p1_2023}.
Practical realization of CLSs is a subject of the running
Horizon Europe EIC-Pathfinder-2021 project TECHNO-CLS \cite{TECHNO-CLS}.
This Project presents a
high-risk/high-gain science-towards-technology breakthrough research
aimed at further development of technologies for crystal samples preparation,
extensive experimental program that includes design and manipulation of
particle beams, detection and characterization
of the radiation, as well as theoretical analysis and advanced
computational modelling for the characterisation of CLS.

In recent years, advanced atomistic computational modeling of the channeling
process and the process of the photon emission as well as of other related
phenomena beyond the continuous potential framework has been carried out using
the multi-purpose computer package \MBNExplorer
\cite{MBNExplorer_2012,MBN_ChannelingPaper_2013} and the special
multi-task software toolkit with graphical user interface
\MBNStudio
\cite{SushkoEtAl_2019-MBNStudio}.
The computation accounts for the interaction of projectiles with separate
atoms of the environments, whereas many different interatomic potentials
implemented in \MBNExplorer support rigorous simulations
of various media \cite{MBNExplorer_Book,MBN-DYSON_Book_2022}.
For simulations of the channeling and related phenomena, an additional
module has been incorporated into \MBNExplorer to compute the motion of
relativistic projectiles along with dynamical simulations of the
environments.
This methodology, called relativistic molecular dynamics (MD), has been
extensively applied to simulate passage of ultra-relativistic charged
particles in oriented crystals accompanied by emission of intensive
radiation.
A comprehensive description of the multiscale all-atom relativistic MD
approach as well as a number
of case studies related to modeling channeling and photon emission by
ultra-relativistic projectiles are presented in
a review article \cite{KorolSushkoSolovyov:EPJD_v75_p107_2021}.

The aim of the present study is to quantify the impact
of the ionising collisions on the channeling efficiency
and on the photon emission spectra for projectile particles
of different charge and energy passing through comparatively thick
crystalline targets.
For doing this we have followed general methodology implemented
in \MBNExplorer to generate particles' trajectories in a
crystalline environment that accounts for randomness in
sampling the incoming projectiles
as well as in displacement of
the lattice atoms from the nodal positions due to thermal
vibrations.
As a result, each trajectory corresponds to a unique crystalline
environment.
Another phenomenon that contributes to the statistical independence
of the simulated trajectories, concerns the events of
inelastic scattering of a projectile particle from the crystal
atoms resulting in the excitation or ionisation of the atom.
These collisions lead to a random change in the direction of the
particle's velocity, which in turn leads to a change  in its transverse
energy.
These quantum events are random and occur on the atomic scale in terms
of time and space, therefore they can be incorporated into the classical
mechanics framework in accordance with their probabilities.

The methodology implemented in \MBNExplorer to account for the
ionising collisions is discussed in Section \ref{Methodology}.
In Section \ref{CaseStudies} we present and discuss the results
of simulations of the channeling process of the sub-GeV
electrons and positrons in different crystalline targets.
A summary of the results obtained is given in Section
\ref{Conclusions}.

\section{Methodology \label{Methodology}}

The \MBNExplorer package
\cite{MBNExplorer_2012,MBNExplorer_Book} implements the method of
relativistic classical MD \cite{MBN_ChannelingPaper_2013}
to model the motion an ultra-relativistic particle
in  an external field or/and in an atomic environment.
The code integrates the relativistic equations of motion of
a particle of mass $m$, charge $q$ and energy $\E$
written in the following form (see, e.g., \cite{Landau2}):
\begin{eqnarray}
\left\{\begin{array}{l}
\dot{\bfv}
= \displaystyle
{
\bfF -  \bm{\beta} \left(\bfF \cdot \bm{\beta}\right) \over m \gamma}
\\
\dot{\bfr} = \textbf{v}
\\
\end{array} \right. \ ,
\label{Methodology:eq.01}
\end{eqnarray}
where $\bfr=\bfr(t)$ and $\bfv=\bfv(t)$ are instantaneous
position vector and velocity of the particle,
$\bm{\beta} = \bfv/c$,
$\gamma = \E/mc^2$ is the Lorentz factor, and
$c$ is the speed of light.

In a general case, the force $\bfF$ acting on the projectile
is the sum of two terms:
\begin{eqnarray}
\bfF = \bfF_{\rm em} + \bfF_{\rm rd} \,.
\label{Methodology:eq.02}
\end{eqnarray}
Here $\bfF_{\rm em}$ denotes the Lorentz force due to
(i) electrostatic field $\bfE$ created by atoms in the environment
and/or by external charges, and (ii) external magnetic field
$\bfB$:
\begin{eqnarray}
\bfF_{\rm em}
= q
\left(
\bfE + \bm{\beta}\times\bfB
\right) .
\label{Methodology:eq.03}
\end{eqnarray}
The second term in (\ref{Methodology:eq.02}),
$\bfF_{\rm rd}$, is a radiation damping force that
accounts for the reaction of the emitted radiation on the
motion of the charged particle, see,
e.g. \cite{Landau2,Jackson}.
Its action leads to a gradual decrease of the particle's
energy.
For ultra-relativistic projectiles of very
high energies (tens of GeV and above) this force becomes
quite noticeable, so that it must be accounted for to
properly simulate the motion, see recent experiments
\cite{Wistisen_EtAl:NatComm_v9_p1_2018,
Wistisen_EtAl-PhysRevRes_v1_033014_2019,
Nielsen_EtAl-PhysRevD_v102_052004_2020}.
The methodology that has been developed to incorporate
$\bfF_{\rm rd}$ into the relativistic MD
framework implemented in the \MBNExplorer
package is described in Ref.\cite{SushkoEtAl:NIMB_v535_117_2023}.

Applied to the passage of a charged particle through an atomic environment,
the environment, the system (\ref{Methodology:eq.01}) describes the
classical motion of the particle in an electrostatic field of the
atoms of the medium.
In this case, the two terms on the right-hand side of eq.
(\ref{Methodology:eq.03})
depend on the intensity of the field, which can be calculated as
is calculated as $\bfE =-\bm{\nabla} \phi(\bfr)$
where $\phi({\bfr})$ is the potential of the field
is equal to the sum of the potentials of the individual atoms.
\begin{eqnarray}
\phi(\bfr)=\sum_j \phi_{\rm at}\left(\left|\bfr-\bfr_j\right|\right),
\label{Methodology:eq.04}
\end{eqnarray}
where $\bfr_j$ denotes the position vector of the nucleus of the
$j$th atom.
For a neutral atom, the potential $\phi(\bfr)$ is a rapidly decreasing
function at distances greater than the average atomic radius.
Therefore, at each integration step of eq.
(\ref{Methodology:eq.01}), the sum in (\ref{Methodology:eq.04})
can only include the atoms inside the sphere of a given cut-off
radius $\rho$ with the centre at $\bfr$.
Typically $\rho$ is chosen to be an order of magnitude larger than the average
atomic radius.
The search for such atoms is done using the
linked cell algorithm implemented in the \MBNExplorer package.
As the particle passes through a medium, the environment is generated
dynamically, so that the simulation box follows the moving particle.
For a crystalline environment
the positions of the atoms are generated taking into account
random displacement from the nodes due to thermal vibrations.
More detailed description on the algorithms used to compute the
trajectories at various scales,
including macroscopically large ones, are presented
in Refs.
\cite{MBNExplorer_2012,MBN_ChannelingPaper_2013,MBNExplorer_Book,
KorolSushkoSolovyov:EPJD_v75_p107_2021,SushkoEtAl:NIMB_v535_117_2023}.

The classical equations of motion (\ref{Methodology:eq.01}) do
not take into account for events of inelastic scattering of a projectile
particle from individual atoms, which  lead to the excitation or
ionisation of the atom.
These events result in two subsequent changes in the particle
motion:
\begin{itemize}
\item
A random decrease in the energy of the projectile (ionisation losses).

\item
A random change in the direction of the projectile motion characterised by two
scattering angles, $\theta$ and $\phi$, measured with respect to the instant
velocity, $\bfv$.
\end{itemize}

Rigorous treatment of inelastic collision events can only be achieved
by means of quantum mechanics.
However, given the random, fast and local nature of such events, they can be incorporated into the classical mechanics
framework according to their probabilities.
This approach has been already implemented in \MBNExplorer
within the framework of irradiation-driven MD
\cite{Sushko_IS_AS_FEBID_2016}.
A similar methodology, described below, has been realised in the code to account for the
ionising collisions.
Its implementation followed the scheme
presented in Ref. \cite{Dechan01}.

For an ultra-relativistic particle passing through a medium, the
probability $\d^2 W$ of the relative energy loss
$\d \Delta = \d\E/\E$ over the path $\d s \approx c \d t$
due to ionising collision with quasi-free electrons
can be calculated using the following expression (see, e.g.,
Eqs. (1.7)-(1.7b) in \cite{RossiGreisen_RMP_v13_p240_1941}
and \S  33.2.7 in \cite{ParticleDataGroup2018}):
\begin{equation}
{\d^2 W \over \d \Delta\, \d s }
=
2\pi r_0^2 q^2\,
{n_{\ee}(\bfr) \over \gamma \Delta^2}\, .
\label{Methodology:eq.05}
\end{equation}
Here $r_0=2.818\times 10^{-13}$ is the classical electron
radius and
$n_{\mathrm{e}}(\textbf{r})$ is the local volume density of
electrons.
In the points $\bfr \neq \bfr_j$, i.e. not at the nuclei,
the density is related to the potential (\ref{Methodology:eq.04})
via the Poisson equation:
\begin{eqnarray}
 n_{\mathrm{e}}(\textbf{r})
 =
 {1 \over 4\pi e}
 \bm{\nabla}^2
 \sum_j \phi_{\rm at}\left(\left|\bfr-\bfr_j\right|\right)\,.
\label{Methodology:eq.06}
\end{eqnarray}
In a single collision, the
value of relative energy loss
$\Delta = (\E-\E^{\prime})/\E$, where $\E$ and $\E^{\prime}<\E$
are the particle's energies before and after the collision,
lies within the interval
\begin{eqnarray}
&& \Delta_{\min} \leq \Delta \leq \Delta_{\max},
\label{Methodology:eq.07}\\
&& \Delta_{\min} = I /\varepsilon,
\qquad
\Delta_{\max}
\approx
\left\{
\begin{array}{ll}
\displaystyle{2\gamma \xi \over 1 + 2\gamma \xi} & \mbox{for a heavy particle}
\\
1                                   & \mbox{for a positron} \\
0.5                                 & \mbox{for an electron}
\end{array}
\right.
\label{Methodology:eq.08}
\end{eqnarray}
where $\xi$ is the electron-to-projectile
mass ratio and  $I$ denotes the atomic ionization potential,
which can be estimated as $I=16 Z^{0.9}$ eV within
the Thomas-Fermi model ($Z$ is the atomic number).

As written, the distribution (\ref{Methodology:eq.05})
holds for $\Delta \ll 1$.
However, since a probability of collisions with $\Delta\gtrsim 1$ is
negligibly small, Eq. (\ref{Methodology:eq.05})
can be applied to the whole range of the $\Delta$ values.

At each step  $\Delta t $ of integration of Eqs.
(\ref{Methodology:eq.01}) an ionising collision is treated as a
probabilistic event.
Once it occurs and the value $\Delta$ is determined, one
calculates a round scattering angle $\theta$ measured with respect
to the instant velocity $\bfv(t)$:
\begin{eqnarray}
\cos\theta
=
\sqrt{\gamma + 1\over \gamma - 1}
\sqrt{
1 - \Delta - \gamma^{-1}
\over
1 - \Delta +\gamma^{-1}
}
-
{\Delta  (\xi-1)
\over
\sqrt{\gamma^{2}-1}\sqrt{(1 - \Delta)^2-\gamma^{-2}}
}
\label{Methodology:eq.09}
\end{eqnarray}
The second scattering angle $\phi$ is not restricted by any
kinematic relations, and its value is generated as a uniform random
deviate between $0$ and $2\pi$.

The algorithm of random generation of the $\Delta$ and
$\theta$ values is as follows \cite{Dechan01}.

The two-fold probability (\ref{Methodology:eq.05}) is normalized as
follows
\begin{equation}
\int_0^{L_{\rm ion}}\, \int_{\Delta_{\min}}^{\Delta_{\max}}\,
{\d^2 W \over \d \Delta\, \d \d s }
\,
\d \Delta\, \d s
= 1
\label{Methodology:eq.010}
\end{equation}
where $L_{\rm ion}$ is a spatial path over which the
probability of an ionising collision with an
arbitrary energy transfer is equal to one:
\begin{equation}
L_{\rm ion}^{-1}
=
{2\pi r_0^2 q^2\, n_{\mathrm{e}}(\textbf{r})\over \gamma}\,
{1 \over \Delta_0},
\qquad
\Delta_0 = {\Delta_{\max}\, \Delta_{\min} \over \Delta_{\max} - \Delta_{\min}}
\approx \Delta_{\min}\,.
\label{Methodology:eq.011}
\end{equation}
Therefore, the probability of an ionising collision
with the energy transfer $\Delta$ to occur over the path
$\Delta s=c\Delta t$ is given by
\begin{equation}
\d W
=
{\Delta s \over L_{\rm ion}}\,
P(\Delta)\,\d \Delta \,,
\quad
P(\Delta) = {\Delta_0 \over \Delta^2}
\label{Methodology:eq.012}
\end{equation}
where the factor $P(\Delta)\, \d \Delta$ stands
the normalized probability of the relative energy loss between
$\Delta$ and $\Delta + \d \Delta$.

To simulate the probability of an ionising collision to take place,
a uniform random deviate $x\in [0,1]$ is generated and compared with
 $\Delta s/L_{\rm ion}$.
If $x \leq \Delta s/L_{\rm ion}$ then the collision happens
and another random deviate $\Delta$ is generated and used
in Eq. (\ref{Methodology:eq.09})
to calculate $\theta$.

The values of $\Delta$, $\theta$ and $\phi$ obtained are used further
to modify the velocity and energy of the projectile at the start of
the next integration step.

To account for or to disregard the ionising collisions
the following schemes are implemented in \MBNExplorer:
\begin{itemize}
\item
the local electron density
$n_{\mathrm{e}}(\textbf{r})$ is calculated using
Eq. (\ref{Methodology:eq.06}) where atomic potentials
$\phi_{\rm at}$ are considered within the frameworks of the
Moli\`{e}re \cite{Moliere} or Fernandes Pacios \cite{Pacios:JCompChem_v96_p7249_1992} approximations.

\item
the local density is substituted with the average
electron density in the amorphous medium:
$n_{\mathrm{e}}(\textbf{r}) = Zn_{\rm at}$ where
$n_{\rm at}$ stands for the volume density of atoms in the
amorphous environment,

\item
to disregard the ionising collisions one sets
$n_{\mathrm{e}}(\textbf{r}) \equiv 0$.
\end{itemize}

\section{Case studies\label{CaseStudies}}

The case studies presented below refer to the channeling of
$\E=270, 855$ and 1500 MeV electrons (e$^{-}$) and 530 MeV positrons (e$^{+}$)
in single diamond (C), silicon (Si) and germanium (Ge) crystals oriented,
with respect to the incident beam, along the (110) and (111) planar
directions.
The beams were assumed to be ideally collimated.
The values of crystal thickness $L$ along the beam used in the simulations
are (i) 200, 200, 400 microns for 270, 855, 1500 MeV electrons, respectively,
and (ii) 1000 microns for the positrons.
For electrons, the aforementioned energy range corresponds that
available at the Mainz Mikrotron (MAMI) facility \cite{MAMI}.
A number of channeling experiments with the electron beam
have been carried out at MAMI with various crystal targets
during the last one and a half decade
\cite{BackeEtAl:NIMB_v266_p3835_2008,BackeLauth:NIMB_v355_p24_2015,
MazzolariEtAl:PRL_v112_135503_2014,
BandieraEtAl:PRL_v115_025504_2015,
BackeEtAl:JPConfSer_v438_012017_2013}.
More recently, within the framework of the TECHNO-CLS project
\cite{TECHNO-CLS}, a new 530 MeV positron beam line has been
designed at MAMI \cite{BackeEtAl:EPJD_v76_150_2022} and the first
experimental tests with the positron beam have been carried out
\cite{MazzolariEtAl:arXiv_2404.08459,Backe:NIMA_v1059_168998_2024}.

\subsection{Statistical analysis of the channeling process
\label{CaseStudies:01}}

For each case study approximately $N_0=1.2\times10^4$ of
trajectories have been simulated and analyzed.
Randomness in  the "entry conditions" leads to  diversity in the scattering events for the different projectiles at the crystal entrance.
Consequently, not all particles start their propagation moving in the  the channeling regime.
To quantify this property,  the acceptance $\calA$ can be introduced
as the ratio
of the number $N_{\rm acc}$ of particles accepted into the channeling
regime to the total $N_0$ number of particles:
$\calA = {N_{\rm acc} / N_0}$.

A significant methodological issue pertains to the formulation of a
criterion for distinguishing between channeling and non-channeling regimes
of projectiles' motion.
Depending on the theoretical approach employed to describe the interaction
of a projectile with a crystalline environment, the criterion can be
introduced in various ways.
For instance, within the continuous potential framework, it is
straightforward to define the channeling particles as those with transverse
energies less than the height of the inter-planar (or inter-axial) potential
barrier $\Delta U$.
Within this framework, the acceptance is determined at the entrance to the
crystal and can be defined as the ratio of the number of particles with
transverse energy less than $\Delta U$ to the total number of particles.
Within the framework of relativistic MD, the simulations are based on
resolution of the EM   (\ref{Methodology:eq.01}),
incorporating the interaction of a projectile particle with individual
atoms of the crystal, as would be observed in reality.
The potential experienced by the projectile varies rapidly during its
motion, thereby coupling the transverse and longitudinal degrees of freedom.
It is therefore necessary to propose an alternative criterion for
identifying the channeling segments in the projectile's trajectory.
In the case of planar channeling, it can be postulated that a projectile is
captured in the channeling mode when the sign of its transverse velocity
undergoes a minimum of two changes within a single channel.
\cite{MBN_ChannelingPaper_2013}.

The collisions (both elastic and inelastic ones) with the crystal atoms
lead to a gradual decrease in the number of channeling particles,
which were initially accepted, with the penetration distance.
This dependence can be characterized by the fraction
$f_{\rm ch,0}(z) = N_{\rm ch,0}(z)/N_{\rm acc}$
\cite{KorolSushkoSolovyov:EPJD_v75_p107_2021}.
Here $N_{\rm ch, 0}(z)$ is the number of particles which have been
accepted at the entrance ($z=0$) and channeled to the distance $z$, where
they dechannel.

The non-accepted particles, as well as the dechanneled particles, experience unrestricted over-barrier motion.
However, due to the collisions, they can be captured in the channeling mode
(rechanneled) at some point in the bulk.
With the notation $N_{\rm ch}(z)$ as the total number of channeling
particles at distance $z$, one can consider the fraction
$f_{\rm ch}(z) = N_{\rm ch}(z)/N_{\rm acc}$, which accounts for
the rechanneling process.

The fraction $f_{\rm ch,0}(z)$ can be used to quantify the dechanneling
effect in terms of the \textit{penetration length} $\Lp$, which stands
for the mean length of the channeling segment of the trajectory of an
accepted particle:
\begin{eqnarray}
\Lp = \int_0^L f_{\rm ch,0}(z) \d z\,.
\label{CaseStudies:eq.01}
\end{eqnarray}
As defined, $\Lp$ depends on the crystal length $L$.
However, for sufficiently large values of $L$, for which
$f_{\rm ch,0}(z) \to 0$ so that the contribution of the interval
$[L,\infty]$ to the integral is negligibly small,
Eq. (\ref{CaseStudies:eq.01}) gives the value of $\Lp$ independent of
$L$.

For both electrons and positrons, an analytical fit for $f_{\rm ch,0}(z)$
can be constructed
using parameters calculated within the diffusion theory of
the dechanneling process (see, e.g.,
\cite{Waho_PRB_v14_p4830_1976,BiryukovChesnokovKotovBook,
Backe:EPJD_v76_153_2022}).
Apart from some initial segment $[0,z_0]$, this fraction can be
approximated by an exponentially decaying function:

For both electrons and positrons, an analytical fit can be constructed for
$f_{\rm ch,0}(z)$ using parameters calculated within the diffusion theory
of the dechanneling process (see e.g.,
\cite{Waho_PRB_v14_p4830_1976,BiryukovChesnokovKotovBook,
Backe:EPJD_v76_153_2022}).
Apart from some initial segment $[0,z_0]$, this fraction can be
approximated by an exponentially decaying function:
\begin{eqnarray}
f_{\rm ch,0}(z) \approx \widetilde{f}_{\rm ch,0}(z)
 = \ee^{-{z-z_0 \over \widetilde{\Ld}}}
\label{CaseStudies:eq.02}
\end{eqnarray}
where $z_0$ and $\Ld$ are fitting parameters.
It can be shown \cite{KorolSushkoSolovyov:EPJD_v75_p107_2021} that the
quantities $L, $$\Lp$, $z_0$ and $\widetilde{\Ld}$ are related as
follows:
$\Lp = z_0 + \Ld\left(1 - \ee^{-(L-z_0)/\Ld}\right)$.
In the case of a thick crystal, $L\gg \Ld$, one finds
$\Lp \approx z_0 + \Ld$.

Unlike $f_{\rm ch,0}(z)$, the dependence of the
fraction $f_{\rm ch}(z)$ on the penetration distance is sensitive to the
sign of the charge of the projectile.
This will be discussed later in the paper.

\begin{figure} [h]
\centering
\includegraphics[clip,width=14cm]{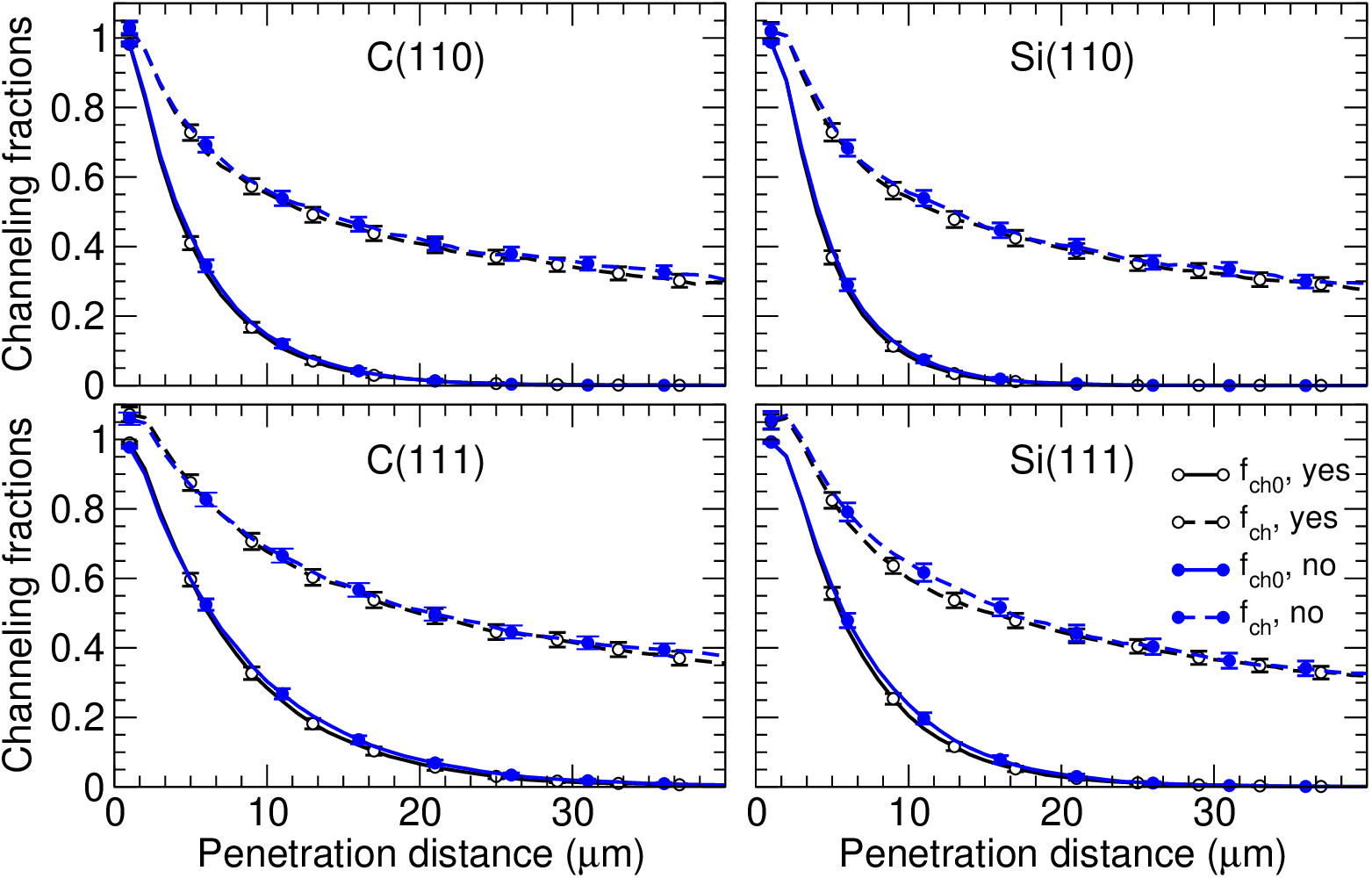}
\caption{
Channeling fractions $f_{{\rm ch}, 0}$ (solid lines) and $f_{\rm ch}$ (dashed lines) versus penetration distance $z$ for 270 MeV electrons in diamond (left column) and silicon (right column) crystals.
The upper row corresponds to the planar channeling along the (110)
direction, the lower row - to the (111) direction.
Curves with open circles show the results obtained taking into account  the ionising collisions (labelled 'yes' in the common legend placed in the upper right graph).
Curves with filled circles represent the dependencies calculated without ionising collisions (labelled 'no' in the legend).
The simulations were performed for the beams of zero divergence.
}
\label{e-270-Fractions.fig}
\end{figure}

\begin{figure} [h]
\centering
\includegraphics[clip,width=14cm]{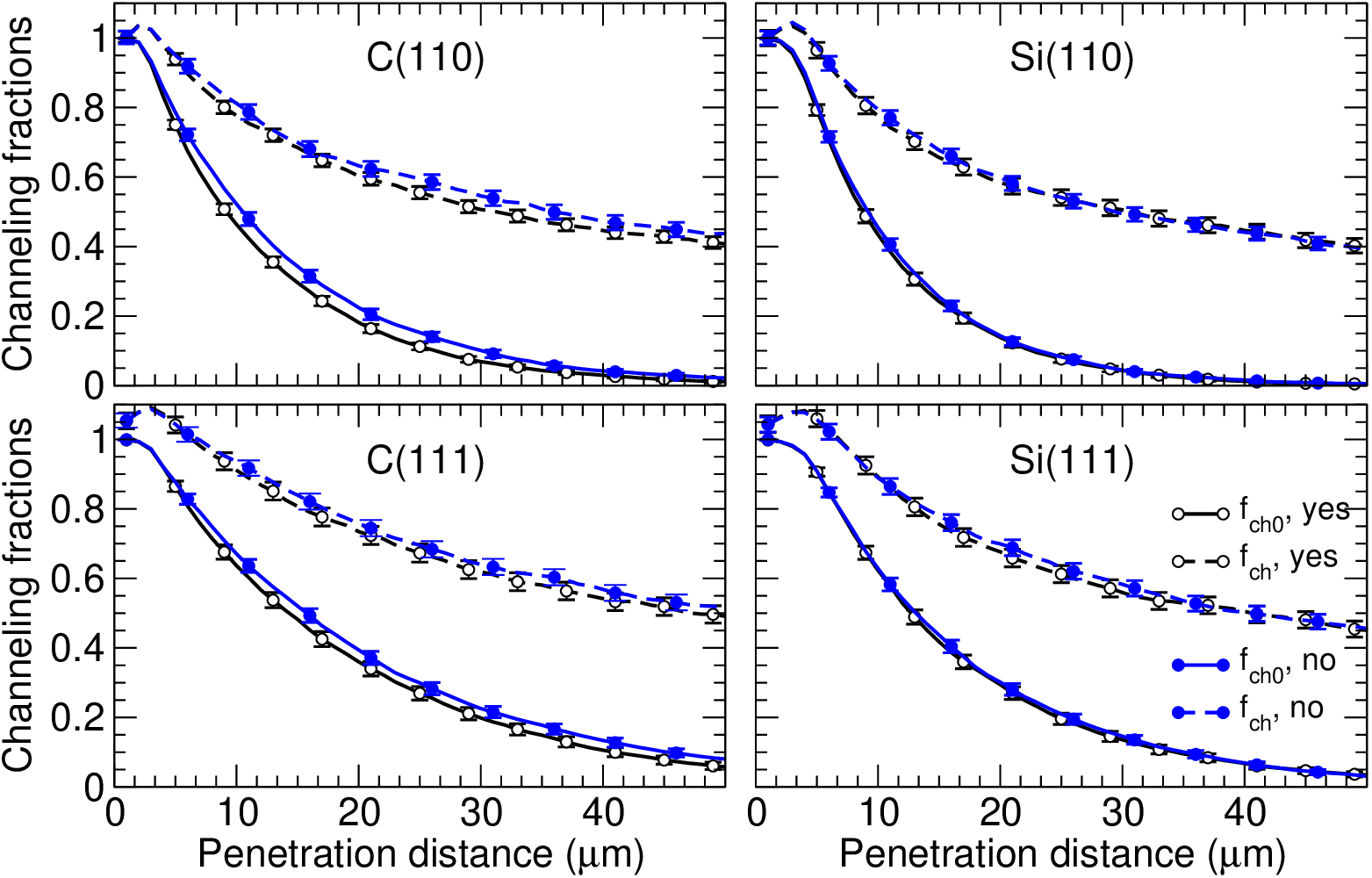}
\caption{
Same as in Fig. \ref{e-270-Fractions.fig} but for
855 MeV electrons.
}
\label{e-855-Fractions.fig}
\end{figure}

\begin{figure} [h]
\centering
\includegraphics[clip,width=14cm]{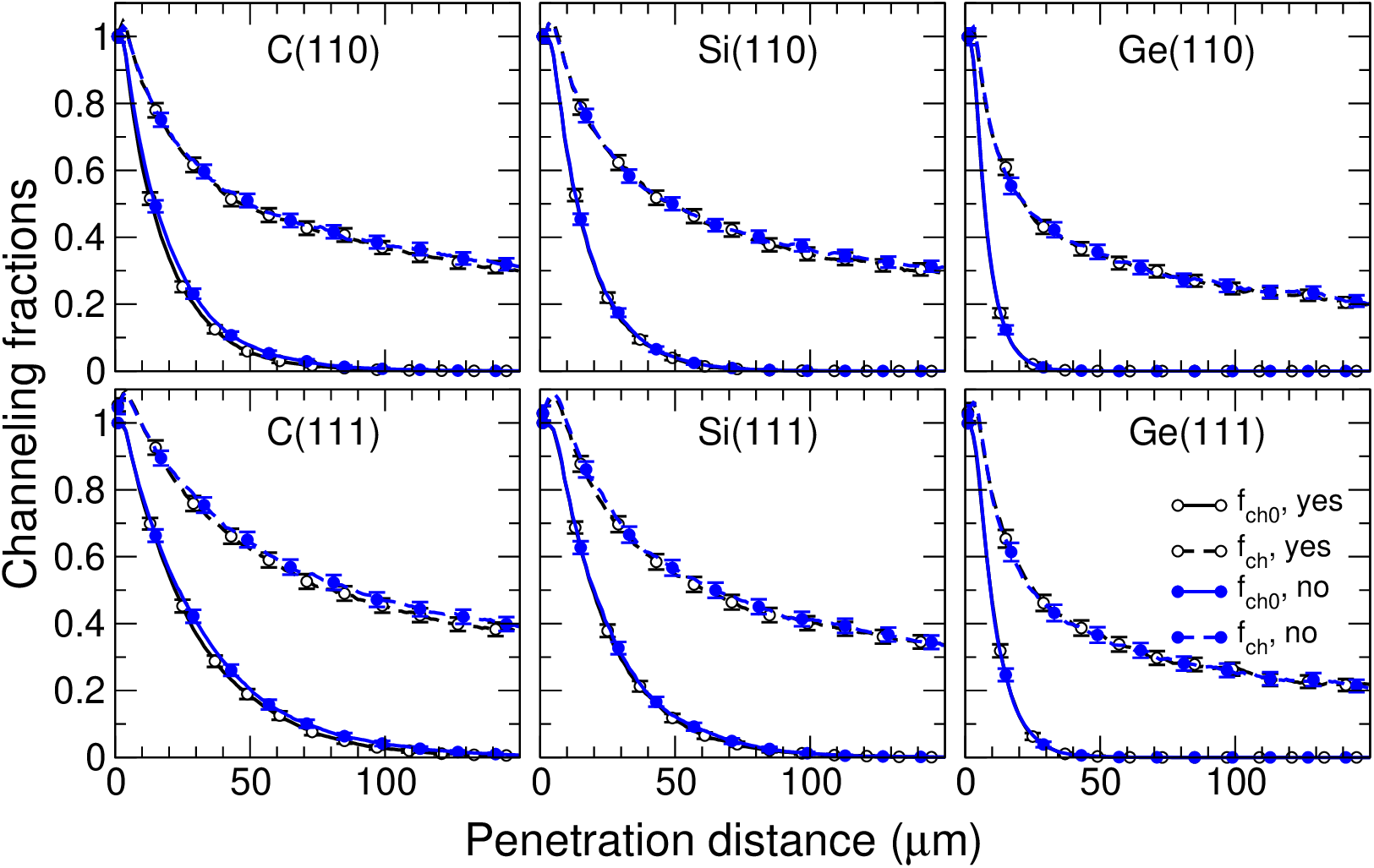}
\caption{
Same as in Fig. \ref{e-270-Fractions.fig} but
(i) for 1500 MeV electrons, and (ii)
adding the case of channeling in the (110) and (111)
channels in germanium crystal, left column.
}
\label{e-1500-Fractions.fig}
\end{figure}

\begin{figure} [h]
\centering
\includegraphics[clip,width=14cm]{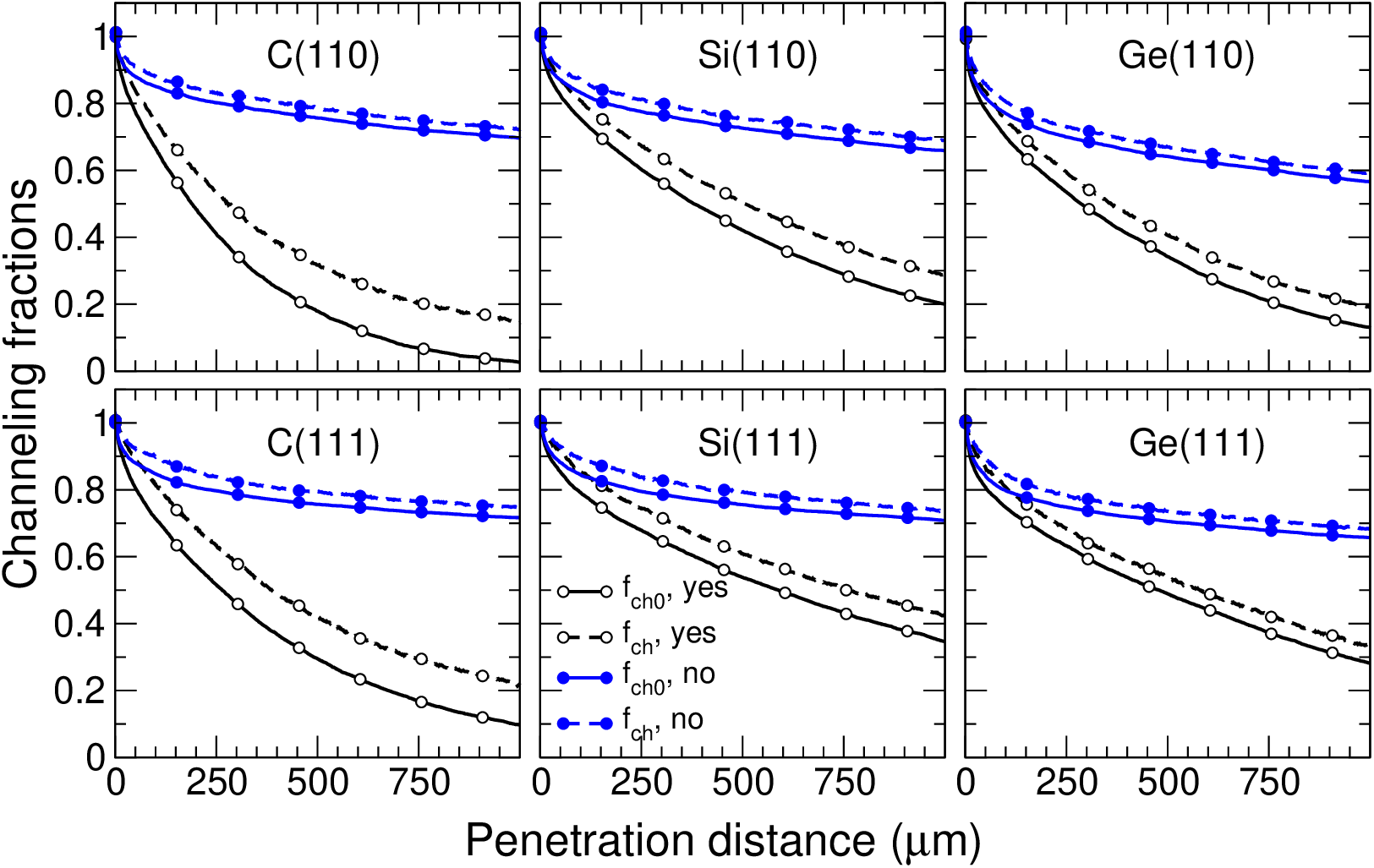}
\caption{
Same as in Fig. \ref{e-1500-Fractions.fig} but
for 530 MeV positrons.This paper presents a quantitative analysis of the impact of inelastic
collisions of ultra-relativistic electrons and positrons, passing through oriented
crystalline targets, on the channeling efficiency and on the intensity  of the
channeling radiation.
The analysis is based on the numerical simulations of the channeling process
performed using the \MBNExplorer software package.
The ionising collisions, being random, fast and local events, are
incorporated into the classical molecular dynamics framework according
to their probabilities.
This methodology is outlined in the paper.
The case studies presented refer to electrons with energy $\E$ ranging from 270
to 1500 MeV and positrons with $\E=530$ MeV incident on thick (up to 1 mm)
single diamond, silicon and germanium crystals oriented along the (110) and
(111) planar directions.
In order to elucidate the role of the ionising collisions,
the simulations were performed with and without account for the
ionising collisions.
The case studies presented demonstrate that both approaches yield highly
similar results for the electrons.
For the positrons, the ionising collisions reduce significantly the
channeling efficiency.
However, it has been observed that this effect does not result in
a corresponding change in the radiation intensity.
The dependencies for the (111) orientations are related to the "wide" part
of the channel. 
}
\label{p-530-Fractions.fig}
\end{figure}

Figures \ref{e-270-Fractions.fig}-\ref{e-1500-Fractions.fig} and
\ref{p-530-Fractions.fig} show the dependence of the channeling
fractions $f_{\rm ch,0}(z)$ (solid curves) and $f_{\rm ch}(z)$
(dashed curves) for electrons and positrons, respectively.
The error bars indicate the statistical uncertainties due to the
finite number of the simulated trajectories.
In each graph, the fractions calculated with account for ionising
collisions (marked 'yes') are compared with those calculated without the
ionising collisions.

Let us first discuss the electron channeling,
Figs. \ref{e-270-Fractions.fig}-\ref{e-1500-Fractions.fig}.
The main feature seen in all graphs is that within the statistical errors
the 'yes' and 'no' dependencies are virtually the same.
The results presented indicate that for electrons
the ionising collisions can be disregarded in the quantitative
description of the dechanneling and rechanneling processes.
This can be understood recalling that negatively charged particles
experience channeling motion in the vicinity of the atomic planes.
Therefore, they experience rather close collisions (i.e. with
comparatively small impact parameter) with the crystal atoms.
For these impact parameters, the change in the transverse momentum of the
particle is mainly due to elastic scattering from the static atomic potential
rather than inelastic scattering from the atomic electrons.
The decrease in transverse momentum can cause the particle to be
re-channeled and thus contribute to the channeling fraction $f_{\rm ch}(z)$.

As the penetration distance increases, the function $f_{\rm ch}(z)$ decreases
much more slowly than the fraction $f_{\rm ch,0}(z)$.
In Ref. \cite{KorolSushkoSolovyov:EPJD_v75_p107_2021} it was shown that
$f_{\rm ch}(z)$ can be well approximated by the following
fitting formula:
\begin{eqnarray}
\widetilde{f_{\rm ch}}(z)
\approx
{\rm erf}\left(\beta\sqrt{\E \over z}\right)
\label{CaseStudies:eq.03}
\end{eqnarray}
where ${\rm erf}(.)$ stands for the error function and $\beta$
is the fitting parameter.
It is mentioned in the cited paper that if $\E$ is measured in MeV
and $z$ in microns then as an initial guess for $\beta$
one can use the value of $(L_{\rm rad}U_0)^{1/2}/106$ where
$L_{\rm rad}$ is the radiation length (in cm) and $U_0$ is the depth of the
interplanar potential (in eV) \cite{KorolSushkoSolovyov:EPJD_v75_p107_2021}.
For sufficiently large distances
${\rm erf}\left(\beta\sqrt{\E/z}\right) \propto z^{-1/2}$.

\begin{figure} [h]
\centering
\includegraphics[clip,width=14cm]{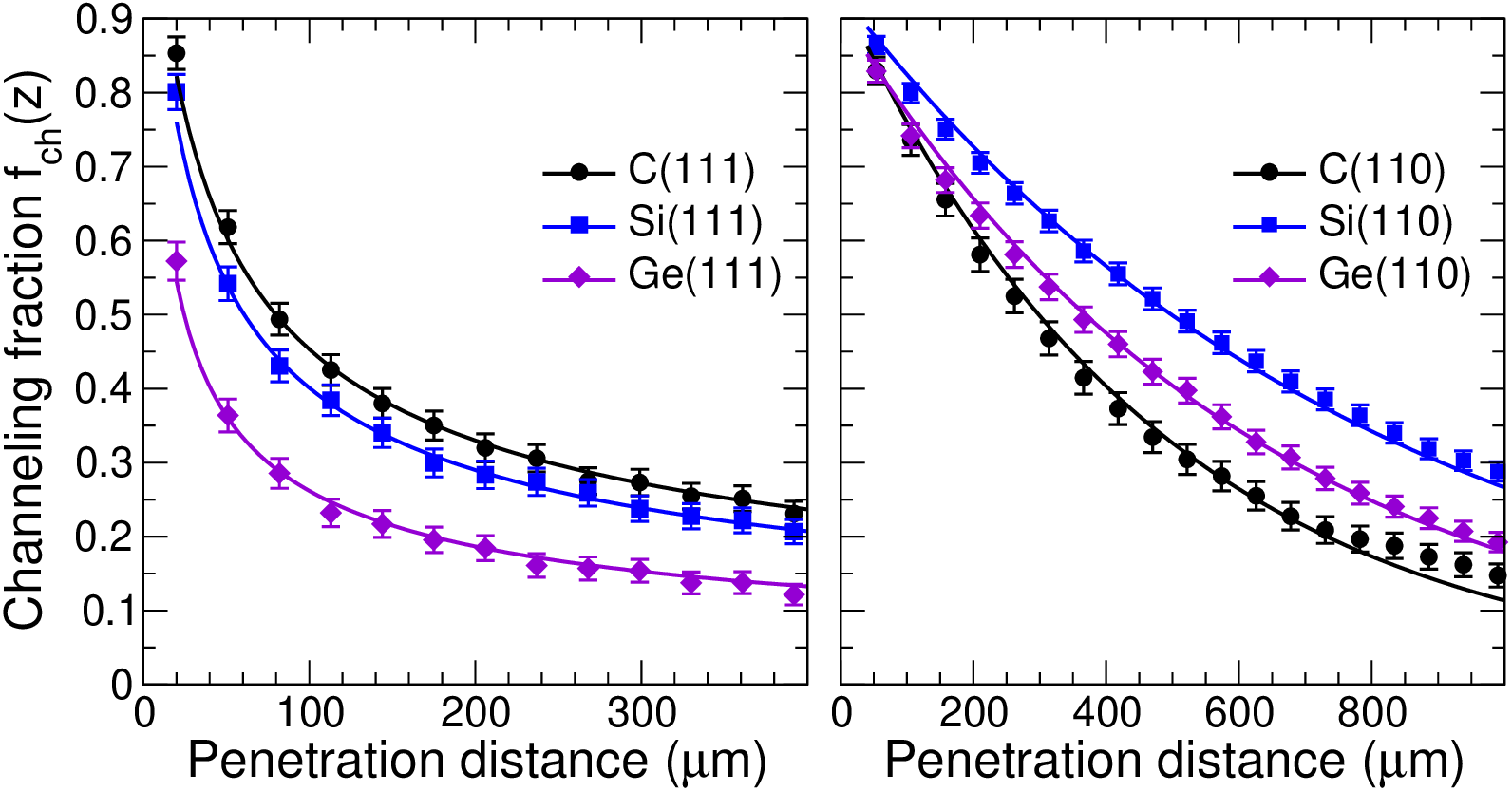}
\caption{
Channeling fraction $f_{\rm ch}(z)$ for $1500$ MeV electrons (left)
and 530 MeV positrons (right).
The crystal orientation is indicated in the legends.
The symbols show the results of the simulations with ionising collisions
taken into account.
The solid curves represent the fitting data calculated using
Eq. (\ref{CaseStudies:eq.03}) for electrons
and the approximation $\widetilde{f_{\rm ch}}(z)\propto \exp(-z/\Lr)$
for positrons.
The values of the fitting parameters $\beta$ and $\Lr$ are given in
Tables \ref{Table.1} and \ref{Table.2}, respectively.
}
\label{e-1500-Fch.fig}
\end{figure}

To illustrate the applicability of (\ref{CaseStudies:eq.03}) we
present Fig. \ref{e-1500-Fch.fig} (left graph), which compares
dependencies $f_{\rm ch}(z)$ obtained from the simulations with
the fitting function $\widetilde{f_{\rm ch}}(z)$.

\begin{turnpage}
\begin{table}[h]
\caption{Acceptance $\calA$, dechanneling length
$\Ld$ (Eq. (\ref{CaseStudies:eq.02})),
penetration length $\Lp$ (Eq. (\ref{CaseStudies:eq.01}))
and fitting parameter $\beta$ (Eq. (\ref{CaseStudies:eq.03}))
calculated for 270, 855 and 1500 MeV electrons incident on
the oriented crystals.
Values of $\Lp$ and $\Ld$ are in microns,
$\beta$ is in $(\mu$m/MeV$)^{1/2}$.
For each channel,
the results obtained with and without taking the ionising collisions
into account are shown in the first and second rows, respectively.
All data refer to the electron beams of zero divergence.
}
\begin{ruledtabular}
\begin{tabular}{ccccccccccccccc}
      &\multicolumn{4}{c}{$\E=270$ MeV}
    & &\multicolumn{4}{c}{$\E=855$ MeV}
    & &\multicolumn{4}{c}{$\E=1.5$ GeV} \\
      &$\calA$&  $\Ld$       &   $\Lp$      &  $\beta$
    & &$\calA$&  $\Ld$       &   $\Lp$      &  $\beta$
    & &$\calA$&  $\Ld$       &   $\Lp$      &  $\beta$\\
\hline
C(110)& 0.73  &$ 4.4\pm0.2$  &$ 5.5\pm0.3$  &$0.104\pm0.006$
    & & 0.74  &$10.5\pm0.4$  &$12.5\pm0.5$  &$0.092\pm0.005$
    & & 0.74  &$16.3\pm0.7$  &$18.9\pm0.9$  &$0.087\pm0.004$ \\
      & 0.74  &$ 4.5\pm0.1$  &$ 5.7\pm0.3$  &
    & & 0.76  &$12.0\pm0.6$  &$14.2\pm0.7$  &
    & & 0.76  &$17.6\pm0.7$  &$20.8\pm1.0$  & \\
\hline
C(111)& 0.68  &$ 7.0\pm0.3$  &$ 8.2\pm0.4$  &$0.125\pm0.005$
    & & 0.70  &$17.0\pm0.8$  &$19.3\pm1.0$  &$0.111\pm0.004$
    & & 0.71  &$27.4\pm1.2$  &$30.5\pm1.5$  &$0.110\pm0.004$ \\
      & 0.69  &$ 7.3\pm0.2$  &$ 8.6\pm0.4$  &
    & & 0.71  &$18.6\pm0.8$  &$21.2\pm1.1$  &
    & & 0.72  &$29.0\pm1.2$  &$32.7\pm1.7$  &   \\
\hline
Si(110)& 0.64 &$ 3.5\pm0.2$  &$ 5.0\pm0.2$  &$0.097\pm0.005$
    & & 0.66  &$ 8.5\pm0.4$  &$11.4\pm0.5$  &$0.090\pm0.004$
    & & 0.74  &$13.8\pm0.6$  &$17.9\pm0.8$  &$0.085\pm0.003$ \\
      & 0.64  &$ 3.7\pm0.2$  &$ 5.2\pm0.2$  &
    & & 0.67  &$ 8.7\pm0.4$  &$11.8\pm0.5$  &
    & & 0.76  &$14.1\pm0.5$  &$18.2\pm0.7$  & \\
\hline
Si(111)& 0.62 &$ 5.1\pm0.2$  &$ 7.1\pm0.3$  &$0.111\pm0.006$
    & & 0.65  &$13.2\pm0.6$  &$16.9\pm0.8$  &$0.102\pm0.004$
    & & 0.66  &$20.3\pm1.0$  &$25.7\pm1.2$  &$0.096\pm0.004$ \\
      & 0.62  &$ 5.5\pm0.3$  &$ 7.4\pm0.3$  &
    & & 0.65  &$13.6\pm0.6$  &$17.3\pm0.8$  &
    & & 0.67  &$21.1\pm1.2$  &$26.5\pm1.3$  &  \\
\hline
Ge(110)&      &              &              &
    & &       &              &              &
    & & 0.59  &$ 5.7\pm0.4$  &$ 8.7\pm0.3$  &$0.057\pm0.003$ \\
      &       &              &              &
    & &       &              &              &
    & & 0.59  &$ 5.8\pm0.2$  &$ 8.8\pm0.4$  & \\
\hline
Ge(111)&      &              &              &
    & &       &              &              &
    & & 0.57  &$ 7.9\pm0.8$  &$11.5\pm0.5$  &$0.061\pm0.003$ \\
      &       &              &              &
    & &       &              &              &
    & & 0.58  &$ 8.0\pm0.8$  &$11.6\pm0.5$  & \\
\end{tabular}
\end{ruledtabular}
\label{Table.1}
\end{table}
\end{turnpage}

Table  \ref{Table.1} summarises the values of the aforementioned
parameters namely the acceptance $\calA$, the penetration distance $\Lp$,
the asymptotic dechanneling length $\Ld$ and the fitting parameter $\beta$
calculated for $\E=270$, 855 and 1500 MeV electrons in different planar
channels.
For each channel shown,
the results obtained with and without taking the ionising collisions
into account are shown in the first and second rows, respectively.

The results of the analysis of the channeling phenomenon for 530 MeV
positrons are presented in Figs. \ref{p-530-Fractions.fig} and
\ref{e-1500-Fch.fig} (right graph), and Table \ref{Table.2}.
In the figures, the curves shown for the (111) orientation of the crystals
refer to the "wide" part of the (111) channel, which exhibits much higher
channeling efficiency than the "narrow" part (see illustrative data in
Appendix \ref{111Pot}.

As demonstrated in Fig. \ref{p-530-Fractions.fig}, the dependencies
$f_{\rm ch,0}(z)$ and $f_{\rm ch}(z)$ exhibit two features that diverge
distinctly from the electron channeling case.

Firstly, it is evident that accounting for the ionising collisions results
in a much steeper decrease of the fractions with the penetration distance.
This is not surprising since positrons effectively channel in the spatial
regions between atomic planes.
Therefore, statistically, they  experience much more distant collisions
with atoms than channeling electrons.
At large impact parameters, the inelastic channels dominate in the total
cross section of the collision between an ultra-relativistic projectile and
an atom.
Consequently, the increase in the transverse momentum occurs at a higher
rate, leading to a faster decrease  of the fractions.
It should be mentioned, nevertheless, that dechanneling due to the elastic
collisions is not at all negligible.
A comparison of the dependencies, calculated with and without
ionisation losses, demonstrates that the elastic scattering provides from 30 \%
(in the case of diamond) up to 50 \% (for Si(111) and Ge(110)) of the total
decrease in the channeling fractions
$f_{\rm ch,0}(z)$ and $f_{\rm ch}(z)$ at the crystal exit.

The second feature to be mentioned is that the replenishment of the channeling
fraction due to the rechanneling phenomenon is much less efficient than
in the case of electron channeling.
Consequently, the dependencies $f_{\rm ch}(z)$ decrease much faster.
In the range of penetration distances $z\leq 1000$ $\mu$m,
considered in the simulations, the behaviour of these dependencies can be
approximated  as an exponentially decaying function $\propto \exp(-z/\Lr)$,
i.e. the same as for $f_{\rm ch,0}(z)$, Eq. (\ref{CaseStudies:eq.02}), but
with different fitting parameter $\Lr$.
Figure \ref{e-1500-Fch.fig}\textit{right} compares the dependencies
$f_{\rm ch}(z)$ obtained from the simulations with the fitting
function.
The values of $\Lr$ are shown in Table \ref{Table.2}
together with the acceptance $\calA$ and the asymptotic dechanneling
length $\Ld$.

\begin{table}[h]
\caption{Acceptance $\calA$, dechanneling length
$\Ld$ and the parameter $\Lr$
(both in microns)
calculated 530 MeV positrons with taking the ionising collisions
into account.
For  the (111) orientation, the $\calA$ and $\Ld$ values indicated in the
upper row pertain to the "wide" part of the channel, while the lower row
corresponds to the "narrow" part.
The values of $\Lr$ refer to the "wide" part only.
The data refer to the positron beam of zero divergence.
}
\begin{ruledtabular}
\begin{tabular}{ccccccc}
       & C(110)   & C(111)   & Si(110)  & Si(111)  & Ge(110)  & Ge(111) \\
\hline
$\calA$& 0.94     & 0.78     &  0.93    & 0.75     & 0.92     & 0.74 \\
       &          & 0.17     &          & 0.18     &          & 0.17 \\
$\Ld$  &$305\pm15$&$445\pm15$&$640\pm15$&$1060\pm30$&$520\pm10$&$900\pm25$\\
       &          &$30\pm 3$&          &$44\pm 4$   &          &$27\pm 2$  \\
$\Lr$  &$470\pm10$&$620\pm15$&$800\pm20$&$1230\pm25$&$615\pm15$&$990\pm20$ \\
\end{tabular}
\end{ruledtabular}
\label{Table.2}
\end{table}

\subsection{Spectral distribution of emitted radiation \label{Spectra}}

In this Section, we present several case studies of the photon
emission spectra by 855 and 1500 MeV electrons and 530 MeV
positrons.
A more extensive presentation and detailed analysis will be
published elsewhere.

For each simulated trajectory, the spectral-angular distribution
$\d^3 E/\d (\hbar\om)\, \d\Om$ of the radiation of energy
$\hbar\om$ emitted within
the solid angle $\d\Om\approx \theta \d\theta\d\phi$
(where $\theta$ and $\phi$ are the polar angles of the emission
direction) is calculated numerically following the algorithm
outlined in Ref. \cite{MBN_ChannelingPaper_2013}.
The spectral distribution of energy radiated within the cone
$\theta_0\ll1$ along the incident beam and averaged over
all trajectories is calculated using the following formula:
\begin{eqnarray}
{\d E(\theta\leq\theta_{0}) \over \d (\hbar\om)}
=
{1 \over N_0}
\sum_{n=1}^{N_0}
\int\limits_{0}^{2\pi}
\d \phi
\int\limits_{0}^{\theta_{0}}
\theta \d\theta\,
{\d^3 E_n \over \d (\hbar\om)\, \d\Om}.
\label{Spectra:eq.01}
\end{eqnarray}
The sum is carried out over all simulated trajectories
of the total number $N_0$; as a result it takes into account the
radiation emitted from the channeling segments of the trajectories
as well as from those corresponding to the non-channeling regime.

\begin{figure} [h]
\centering
\includegraphics[clip,width=14cm]{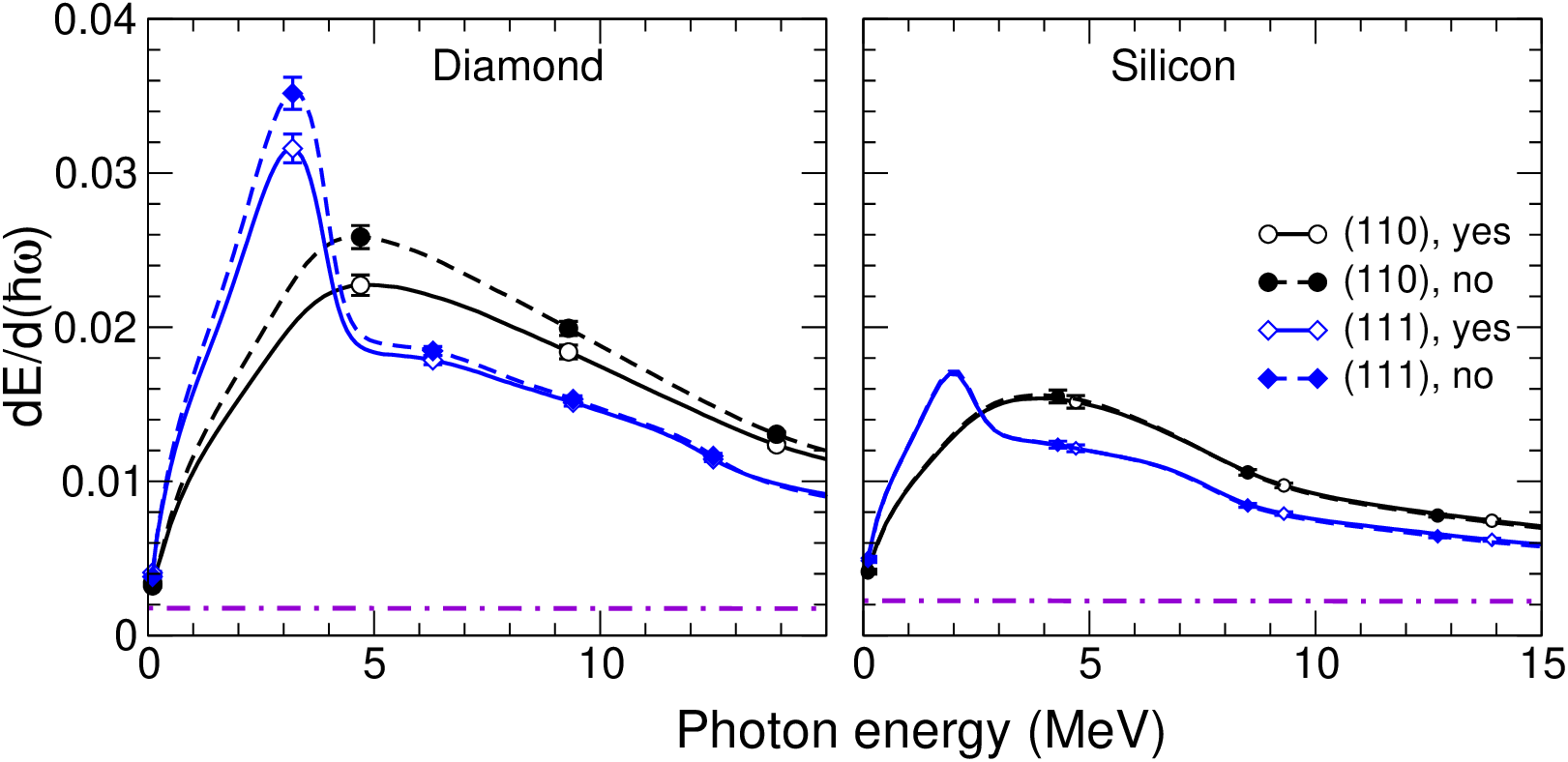}\\ \vspace*{0.5cm}
\includegraphics[clip,width=14cm]{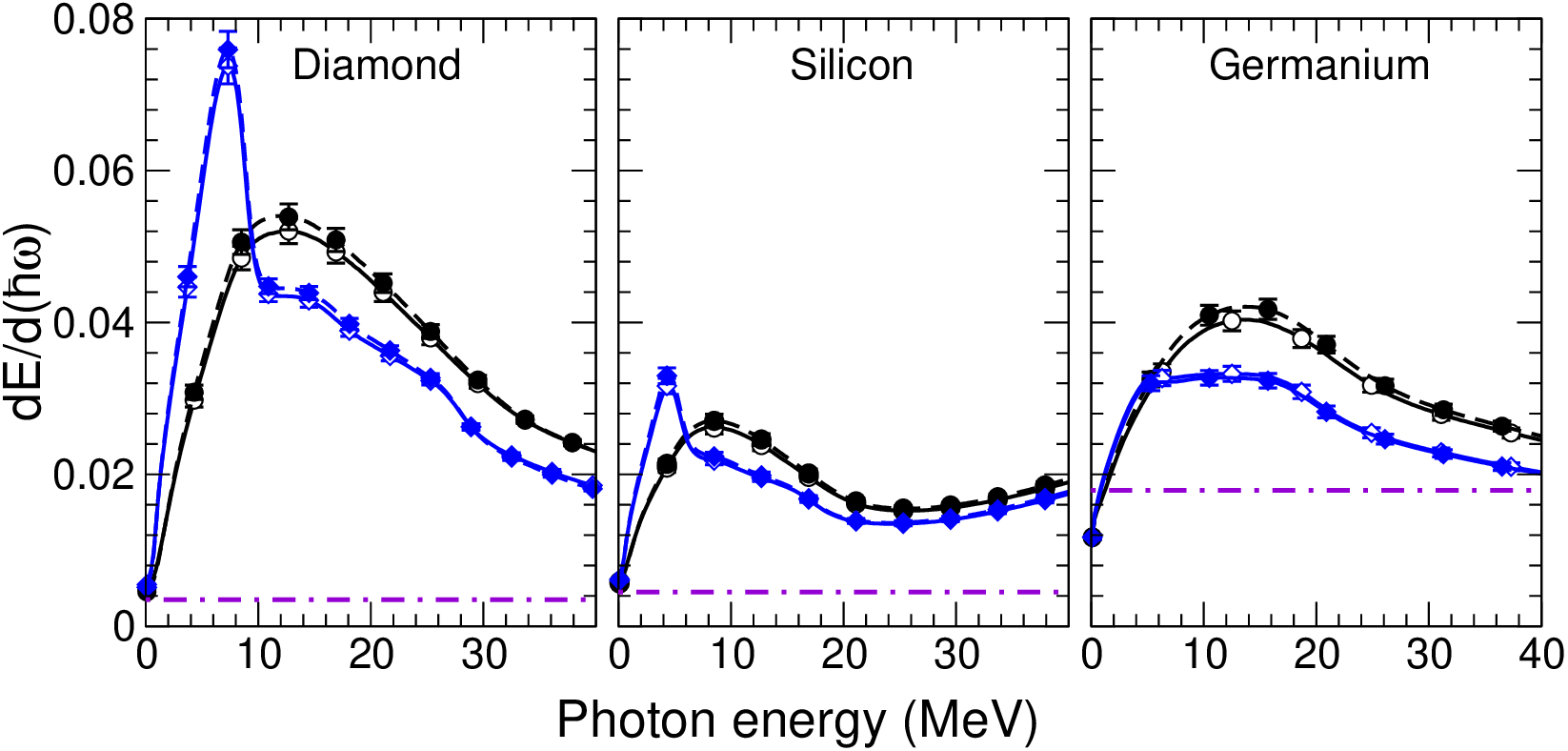}
\caption{
Spectral distribution of the radiation emitted by 855 MeV (upper row) and
1500 MeV (lower row) electrons channeled in (110) and (111) channels in
the diamond, silicon and germanium crystals, as indicated.
The solid curves show the results obtained taking into account the ionising
collisions (labelled 'yes' in the common legends placed in the
upper right graph).
The dashed curves represent the dependencies calculated without ionising
collisions (labelled 'no' in the legend).
The spectra shown correspond to the crystal thicknesses
$L=200$ and 400 $\mu$m and the emission cones
$\theta_0=3$ and 1 mrad for the 855 and 1500 MeV projectiles, respectively.
The dashed-dotted lines show the spectrum of the incoherent
bremsstrahlung emitted in the amorphous target of the same thickness.
}
\label{e-855-1500-Spectra.fig}
\end{figure}

The spectral distributions of radiation emitted by electrons are
presented in  Fig. \ref{e-855-1500-Spectra.fig}.
The upper row corresponds to 855 MeV electrons propagating through
crystalline targets with a thickness of 200 $\mu$m, while
the distributions for 1500 MeV electrons in 400 $\mu$m thick crystals
are shown in the graphs in the lower row.
For the lower energy, the spectra were computed for the cone with
$\theta_0=3$ mrad, which exceeds the natural emission cone
$1/\gamma$ by a factor of approximately five.
For the higher energy, the cone considered 3 mrad is
approximately $3/\gamma$.
The common legend placed in the top-right graph indicates the crystals'
orientation as well as marks the distributions calculated with account
for ionising collisions ('yes') and without the
ionising collisions ('no').
For the sake of comparison, the dashed-dotted line in each graph
shows the spectrum of bremsstrahlung emitted in the amorphous medium of
the same thickness.
These spectra were calculated within the Bethe–Heitler
approximation \cite{BetheHeitler1934,BetheMaximon1954}.

For both energies of the electron beam the distributions are calculated
in the photon energy ranges where the spectrum is
dominated by the channeling radiation \cite{ChRad:Kumakhov1976}.
For the purpose of this paper it is important to note that, similar to
the channeling fractions, the account for the ionising collision does
not affect the spectral distributions.
The most significant discrepancy (of approximately 10 per cent in the
maximum of the spectra) between the 'yes' and 'no' curves is observed
for  855 MeV electrons in the diamond targets.

\begin{figure} [h]
\centering
\includegraphics[clip,width=14cm]{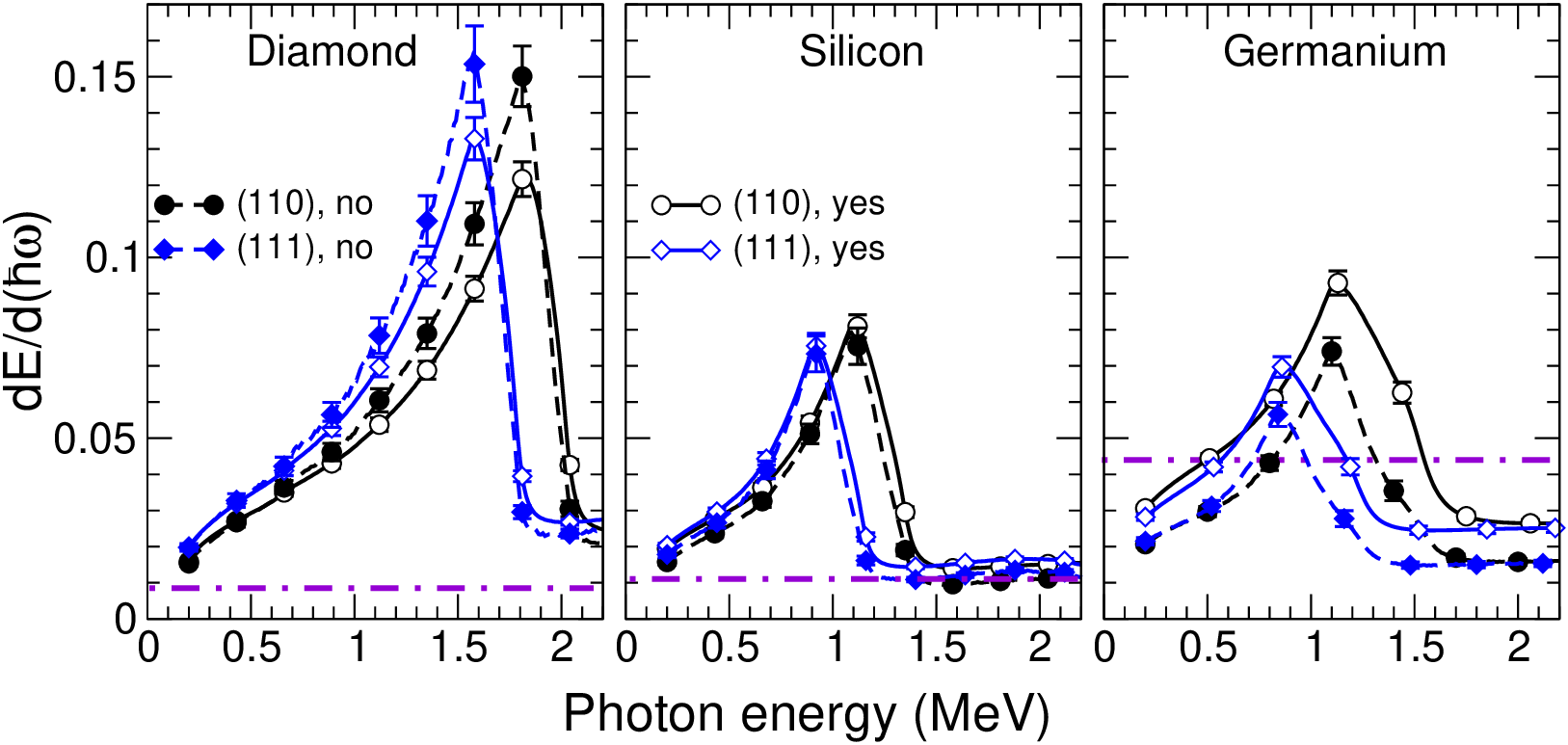}
\caption{
Same as in Fig. \ref{e-855-1500-Spectra.fig} but
for 530 MeV positrons.
In all cases the crystal thicknesses is $L=1000$ $\mu$m and
the emission cone $\theta_0=1$ mrad.
The dashed-dotted lines show the spectrum of the incoherent
bremsstrahlung calculated in the Bethe-Heitler
approximation.
}
\label{p-530-Spectra.fig}
\end{figure}

The results for the spectral distributions of radiation by 530
MeV positrons in $L=1000$ $\mu$m-thick diamond, silicon and germanium
oriented crystals are presented in Fig. \ref{p-530-Spectra.fig}.
In all cases the emission cone is
$\theta_0=1\, \mbox{mrad} \approx 3/\gamma$.

Considering a significant decline in channeling efficiency due to  ionising
collisions, see Fig. \ref{p-530-Fractions.fig}, one could expect to have a similar
impact on the spectral distribution of the channeling radiation.
It is seen that when ionising collisions are taken into account (the solid curves
marked 'yes' in the legend), the peak value of the channeling radiation decreases
by approximately 20-25 per cent for diamond and germanium, and remains virtually
unchanged for silicon crystal.
A possible explanation for this feature is based on the assumption that, although
ionising collisions lead to a reduction in the average length of the channeling
segments they also lead to a gradual increase in the amplitude of the channeling
oscillations for the channeling particle.
The former phenomenon tends to reduce the intensity of the channeling radiation,
while the latter has the opposite effect.
A more detailed quantitative analysis of this assumption is currently on the way.

\section{Conclusions \label{Conclusions}}

A quantitative analysis of the impact of inelastic
collisions on the channeling efficiency and  the radiation emission by
270-1500 MeV electrons  and 530 MeV positrons incident on diamond,
silicon and germanium single crystals oriented along the (110) and
(111) planar directions has been reported.
The crystals thickness along the incident beam was 200-400 $\mu$m
for the electrons and 1000 $\mu$m for the positrons.

The simulations were performed by means of the \MBNExplorer software
package within the framework of classical relativistic MD.
The incorporation of inelastic scattering events of projectile particles
from crystal atoms into the classical framework was conducted according
to their probabilities.

In order to elucidate the role of the ionising collisions,
calculations were performed of the channeling fractions versus the
penetration distance and of the spectral distribution of the emitted
radiation, with and without account for these collisions.

The case studies presented for the electron beams have demonstrated
that both approaches lead to very close results.
The largest impact of ionising collisions on the values of the
channeling fractions and of the channeling radiation intensity, seen for the
diamond targets, is on the level of ten per cent.
This discrepancy is significantly reduced in the heavier crystals.
A comparatively weak role for ionising collisions can be explained
by recalling that electrons channel in the vicinity of
atomic planes.
In this regime, the change in the transverse energy of the
particle is mainly due to elastic scattering from the static atomic potential
rather than inelastic scattering from atomic electrons.

The situation is less straightforward for positrons.
They channel into the spatial regions between the atomic levels and therefore experience more distant collisions in which the inelastic channels dominate
in the total cross section of a scattering from an atom.
The simulations of the channeling motion have supported this conclusion
leading to a much faster decrease in the channeling fractions with
increasing penetration distance.
The dechanneling rate due to inelastic collisions increases,
ranging from 50 per cent (for Si(111) and Ge(110)) up to 70 per cent
(for diamond) of the total dechanneling rate.
However, the increase in the dechanneling rate does not lead to a
comparable change in the radiation emission spectra.
This phenomenon may be explained by the assumption that ionising collisions reduce
the average length of the channeling segments, but increase the amplitude of the
channeling oscillations. The former reduces the intensity of the channeling
radiation, while the latter increases it.
A more detailed quantitative analysis of the particle dynamics and a more
comprehensive presentation of the data on the radiation emission spectra will be
presented elsewhere.

\section*{Acknowledgements}

We acknowledge support by the European Commission
through the N-LIGHT Project within
the H2020-MSCA-RISE-2019 call (GA 872196)
and the EIC Pathfinder Project TECHNO-CLS
(Project No. 101046458).
We also acknowledge the Frankfurt Center for Scientific
Computing (CSC) for providing computer facilities.

\appendix

\section{The (111) interplanar potentials in diamond, silicon and
germanium crystals
\label{111Pot}}

The results presented in the main text have been obtained using the
all-atom relativistic MD approach implemented in the
\textsc{MBN Explorer} software package
\cite{MBNExplorer_2012,MBN_ChannelingPaper_2013,MBNExplorer_Book,%
MBN_ChannelingPaper_2013}.
This approach goes beyond the continuous potential model \cite{Lindhard}.
Therefore, the numerical data presented below in this section are for
illustrative purposes only.

\begin{figure} [h]
\centering
\includegraphics[clip,width=\textwidth]{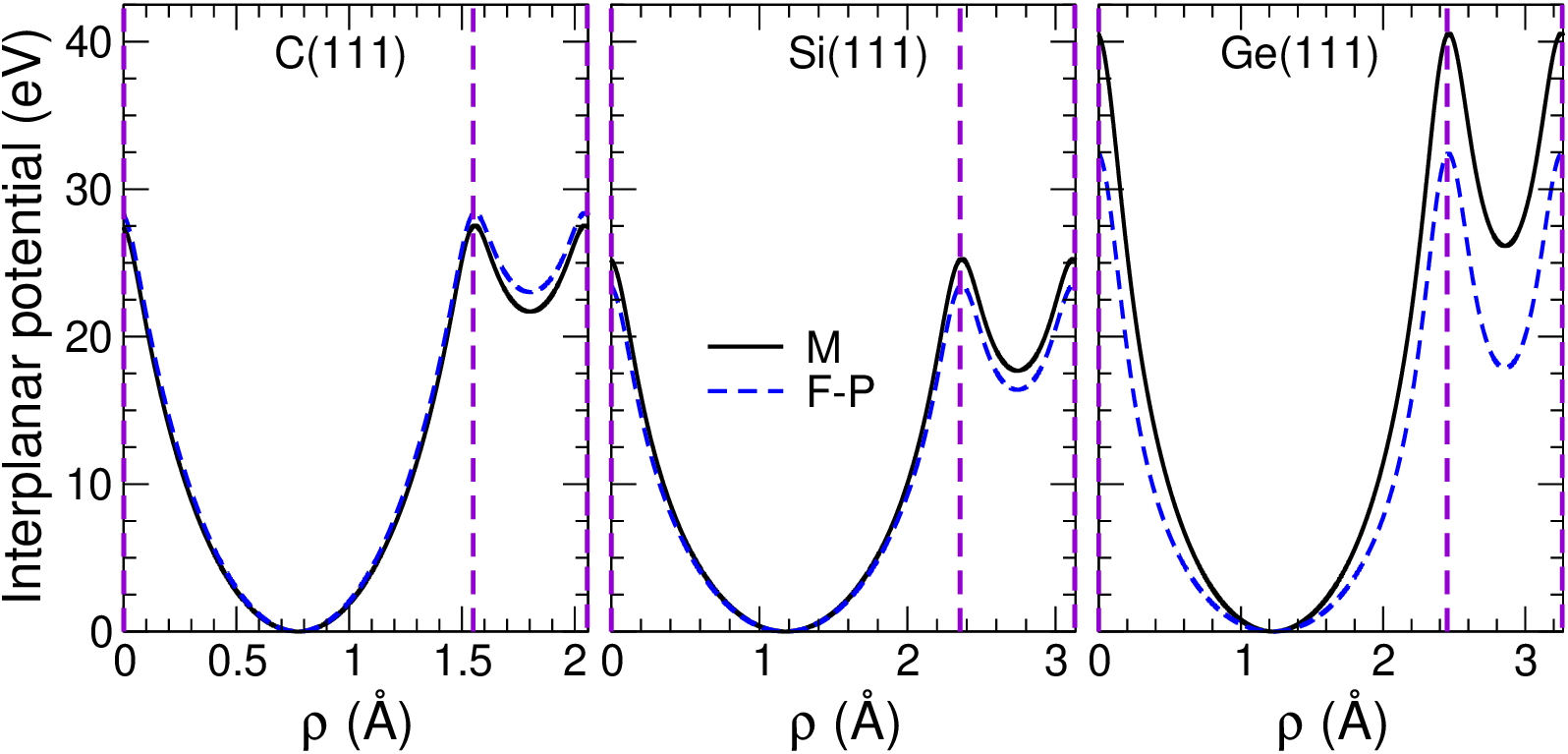}
\caption{
Continuous interplanar potential for a positron in the diamond (C),
silicon and germanium crystals with the (111) orientation.
The vertical dashed lines mark the positions of the crystallographic
(111) planes and illustrate the wide and narrow channels.
The solid and dashed curves represent the potentials
calculated using the Moli\`{e}re (M)  \cite{Moliere}
and Fernandes Pacios (F-P) \cite{Pacios:JCompChem_v96_p7249_1992}
parametrisations of the atomic potentials, respectively.
}
\label{p-111PlanarPot.fig}
\end{figure}

In crystals of the diamond group (diamond itself, silicon, germanium)
the (111) crystallographic planes are arranged with two different
alternating spacings.
Expressed in terms of the lattice constant a, the wide interplanar distance
is $d_{\rm W}=0.75 a/\sqrt{3}$ and the narrow spacing is three times less,
$d_{\rm N}=0.25 a/\sqrt{3}$.
The channeling motion of a positively charged projectile particle occurs in
between two neighbouring planes.
Consequently, in the crystal with the (111) orientation there are wide
and narrow positron channels.

Figure \ref{p-111PlanarPot.fig} illustrates the continuous interplanar
potentials in the wide and narrow (111) positron channels in the diamond, silicon and germanium crystals.
The vertical dashed lines mark the positions of the crystallographic
(111) planes.
In each case, the interplanar potential is calculated by summing the
continuous potentials U of individual planes.
In turn, the potentials U are calculated by averaging out the potentials of
individual atoms, considering their displacement from the nodal positions
due to the thermal vibrations.
The dependencies presented refer to a temperature $T=300$ $^{\circ}$C.
The solid and dashed curves in the figure represent the potentials
calculated using the Moli\`{e}re (M)  \cite{Moliere}
and Fernandes Pacios (F-P) \cite{Pacios:JCompChem_v96_p7249_1992}
parametrisations of the atomic potentials, respectively.
More technical details on the calculation of the continuous planar
potentials can be found in Ref.
\cite{KorolSushkoSolovyov:EPJD_v75_p107_2021}.

It is clear that, in all cases, the potential well in the wide channel is much
deeper.
This, and the big difference in the widths of the channels, indicates that
the positron channeling efficiency of the wide channels is much higher than
that of the narrow channels.
This is also shown by the numerical simulations, which show that the dechanneling
length, $\Ld$ in narrow channels is one order of magnitude less than in wide
channels (see Table \ref{Table.2}).

\section*{References}

\bibliography{references_2025_02_13.bib}

\end{document}